\documentclass[8.5pt,twoside,twocolumn]{article}
\usepackage[super,sort&compress,comma]{natbib}
\usepackage[version=3]{mhchem}
\usepackage{balance}
\usepackage{times,mathptm}
\usepackage{sectsty}
\usepackage{graphicx}
\usepackage{lastpage}
\usepackage[format=plain,justification=justified,singlelinecheck=false,font=small,labelfont=bf,labelsep=space]{caption}
\usepackage{fancyhdr}
\usepackage{xcolor}

\usepackage{fixltx2e}
\usepackage{xspace}
\newcommand{\USERMESO}{\texttt{\textsubscript{\textit{USER}}MESO}\xspace}

\begin{document}

\thispagestyle{plain}
\fancypagestyle{plain}{
\renewcommand{\headrulewidth}{1pt}}
\renewcommand{\thefootnote}{\fnsymbol{footnote}}
\renewcommand\footnoterule{\vspace*{1pt}%
\hrule width 3.4in height 0.4pt \vspace*{5pt}}
\setcounter{secnumdepth}{5}

\makeatletter
\renewcommand\@biblabel[1]{#1}
\renewcommand\@makefntext[1]%
{\noindent\makebox[0pt][r]{\@thefnmark\,}#1}
\makeatother
\renewcommand{\figurename}{\small{Fig.}~}
\sectionfont{\large}
\subsectionfont{\normalsize}

\fancyfoot[RO]{\footnotesize{\sffamily{1--\pageref{LastPage} ~\textbar  \hspace{2pt}\thepage}}}
\fancyfoot[LE]{\footnotesize{\sffamily{\thepage~\textbar\hspace{3.45cm} 1--\pageref{LastPage}}}}
\fancyhead{}
\renewcommand{\headrulewidth}{1pt}
\renewcommand{\footrulewidth}{1pt}
\setlength{\arrayrulewidth}{1pt}
\setlength{\columnsep}{6.5mm}
\setlength\bibsep{1pt}

\newcommand*{\citen}[1]{%
  \begingroup
    \romannumeral-`\x 
    \setcitestyle{numbers}%
    \cite{#1}%
  \endgroup
}

\twocolumn[
  \begin{@twocolumnfalse}
\noindent\LARGE{\textbf{Mesoscale modeling of phase transition dynamics of thermoresponsive polymers$^{\dag}$}}
\vspace{0.4cm}

\noindent\large{\textbf{Zhen Li,\textit{$^{a}$} Yu-Hang Tang,\textit{$^{a}$} Xuejin Li,\textit{$^{a}$} and
George Em Karniadakis$^{\ast}$\textit{$^{a}$}}}\vspace{0.0cm}


 \end{@twocolumnfalse} \vspace{0.4cm}
  ]

\noindent\textbf{We present a non-isothermal mesoscopic model for investigation of the phase transition dynamics of thermoresponsive polymers. Since this model conserves energy in the simulations, it is able to correctly capture not only the transient behavior of polymer precipitation from solvent, but also the energy variation associated with the phase transition process. Simulations provide dynamic details of the thermally induced phase transition and confirm two different mechanisms dominating the phase transition dynamics. A shift of endothermic peak with concentration is observed and the underlying mechanism is explored.}
\section*{}
\vspace{-1cm}
\footnotetext{\textit{$^{a}$~Division of Applied Mathematics, Brown University, Providence, RI 02912, USA. E-mail: george\_karniadakis@brown.edu}.\\
$^{\dagger}$~Electronic supplementary information (ESI) available: Equations, simulation parameters and additional data.}

Thermoresponsive polymers (TRPs) have attracted increasing attention in the last two decades because of their great potential applications in various chemical and biological systems \cite{{2008Klouda},{2013D_Roy}}, i.e., controlled drug delivery, smart materials, bioseparations and filtration. Most applications of TRP have relied on a drastic and discontinuous change of their solubility in given solvents with temperature \cite{{2008Klouda},{2010Betancourt}}. In particular, the temperature-composition diagram of TRP involves a miscibility gap. Depending on the miscibility gap if it appears at low or high temperatures, the critical temperature $T_c$ is known as the lower critical solution temperature (LCST) or the upper critical solution temperature (UCST), respectively. LCST-type TRPs are hydrophilic and highly mixed with the surrounding solvent at low temperatures, but become hydrophobic and precipitate from the solvent above the critical phase transition temperature, while UCST-type TRPs exhibit the opposite behavior \cite{2013D_Roy}. The underlying mechanism of this solubility transition with temperature is related to the role of hydrogen bonds \cite{2009Dybal}. For LCST-type TRP at low temperature $T<T_c$, hydrogen bonds are generated between solvent and polymer molecules. Therefore, the polymers show hydrophilic properties and can be easily dissolved into the solvent. However, when the temperature is increased above the critical temperature $T>T_c$, those solvent-polymer hydrogen bonds are disturbed and polymer-polymer hydrogen bonds dominate the dynamics, which makes the polymer become hydrophobic and precipitate from the solvent.

\begin{figure}[b!]
  \centering
  \includegraphics[width=0.32\textwidth]{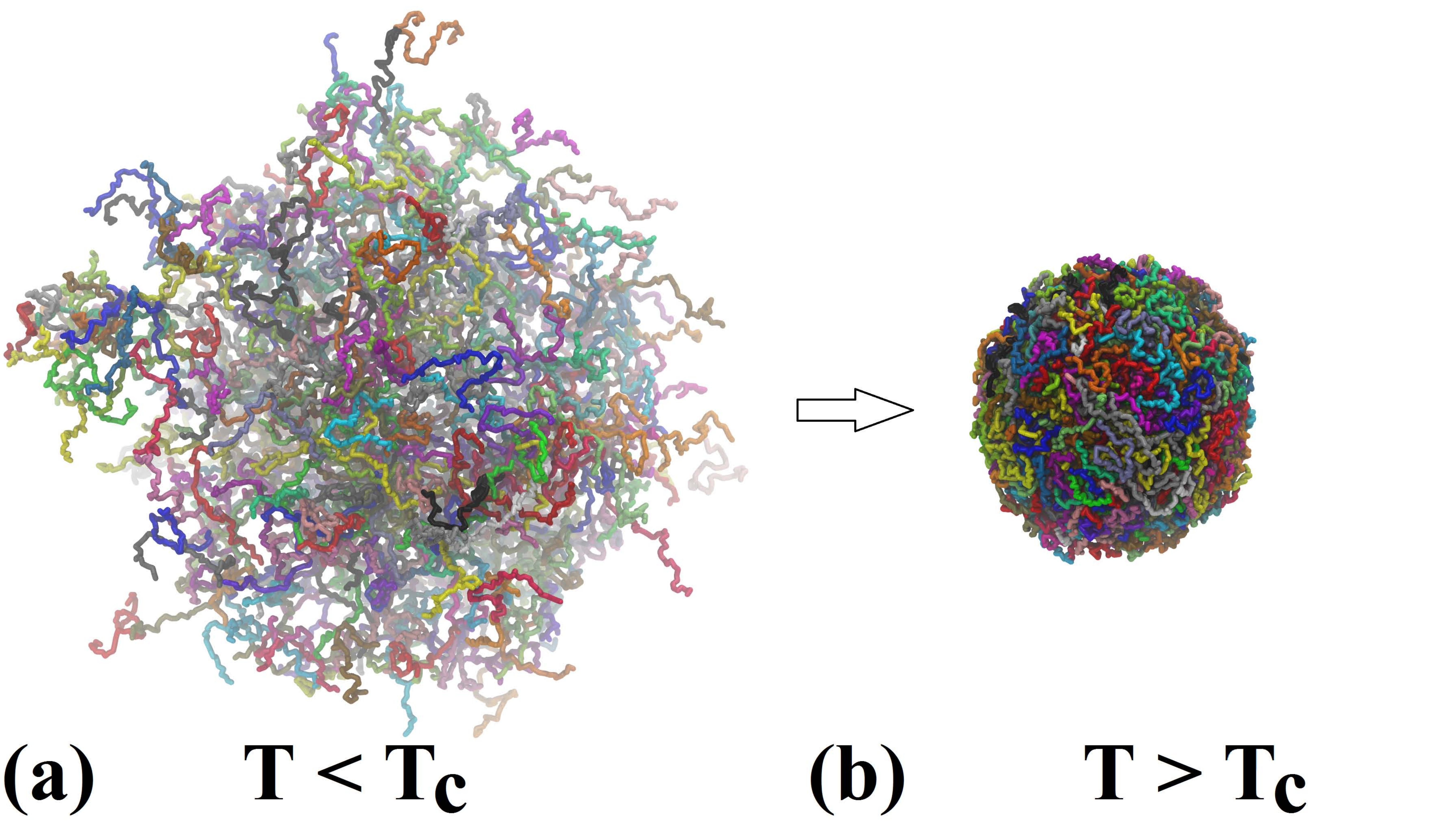}\\
  \caption{A single temperature-sensitive microgel bead consisting of LCST-type TRP chains showing (a) a fully swollen state at low temperature $T<T_c$, and (b) a fully collapsed state at high temperature $T>T_c$. Various polymer chains are visualized with different colors.}\label{fig:1}
\end{figure}

In practical applications of TRP, temperature-sensitive microgels/micelles are often used for the functional element \cite{2009Lyon}. The major building block for these temperature-sensitive microgels is TRP. Among them, poly(N-Isopropylacrylamide) (PNIPAM) is the most investigated material and was extensively used for the construction of temperature-sensitive microgels. Specifically, PNIPAM has a LCST around $32^\circ C$ between room and body temperatures, which makes it a prominent candidate in biomedical applications \cite{2013D_Roy}. As a matter of fact, the applications of TRP highly depend on the evolution of the microstructure of microgels in the phase transition process. Predicting the performance of TRP-based materials requires a deep understanding of the thermally induced phase transition dynamics. Usually, experiments are able to observe the coil-to-globule transition of polymers by measuring the light transmittance rate, and study the static microstructure of microgels using NMR, light scattering and transmission electron microscopy \cite{2012Sun}. It is well-known that the static microstructure of LCST-type microgels is swollen at low temperatures and collapsed at high temperatures, as shown in Fig. \ref{fig:1}. To this end, some theories have been developed for understanding the experimental observations at the molecular level \cite{2003Wu}. However, it is difficult for both experiment and theory to provide dynamic details of the transition process, which is very important for
clarifying the phase transition of TRP. Alternatively, computational simulation techniques are able to provide details of these dynamic processes and can be used to assist $in\ silico$ design for specific applications of TRP.

The typical diameter of a single microgel bead ranges between $50~nm$ and $5~\mu m$ \cite{2000Pelton}, and the time scale for the phase transition of individual PNIPAM-based microgel beads is in the order of $100~ns$ \cite{{2004Reese}}. Simulating a dynamic process lasting for hundreds of nanoseconds with a length scale of several microns is difficult for conventional atomistic methods, but it is in the comfortable temporal and spacial scales of mesoscopic approaches \cite{2013Mills}. Therefore, in the present work, we develop a non-isothermal mesoscopic model based on dissipative particle dynamics (DPD) to investigate the thermally induced phase transition of TRP. DPD is a particle-based mesoscopic approach, which is usually considered as a coarse-grained molecular dynamics (MD) model \cite{2014Li_SM}. The classic DPD method was designed for simulating isothermal hydrodynamics, which is not valid for non-isothermal processes because of the violation of energy conservation \cite{1997Espanol}. To conserve the energy of the system, an extension of DPD was developed by including the mesoscopic energy equation \cite{1997Espanol,2014Z_Li}. The energy-conserving DPD model is known in the literature as eDPD, and it has been demonstrated that eDPD conserves the energy of fluid systems in simulations and can capture the correct temperature-dependent properties of fluids \cite{2014Z_Li}. In this paper, we extend the eDPD framework to modeling the temperature sensitivity of TRP (for details on the eDPD formulations, see ESI$^\dag$).

The potential between TRP and solvent is sensitive to the temperature changes, and the polymer-solvent interaction parameter $\chi$ is a function of temperature \cite{2003Wu}. In the DPD method, the Flory-Huggins $\chi$-parameter is linear with respect to the excess repulsion $\Delta a$ \cite{1997Groot}, which is defined by $\Delta a=a_{sp}-a_{ss}$ where $s$ represents solvent and $p$ stands for polymer. To model the thermally induced phase transition of TRP, we define the excess repulsion $\Delta a$ as a function of temperature to consider the temperature-dependence of the Flory-Huggins $\chi$-parameter. In practice, we take the repulsion parameters between particles as $a_{ss}(T)=a_{pp}(T)=75k_BT/\rho$, and the cross terms $a_{sp}(T)=75k_BT/\rho+A_0+\Delta A/[1.0+exp(-\tau\cdot(T-T_c))]$ containing a sharp change by $\Delta A$ at $T=T_c$ (see Fig. S2, ESI$\dag$). Since the conservative force between particles is given by $\mathbf{F}^C_{ij}=a_{ij}(T)(1-r_{ij}/r_c)\mathbf{e}_{ij}$ and the corresponding potential is $U_{ij}=\frac{1}{2}a_{ij}(T)r_c(1-r_{ij}/r_c)^2$, the pair potential between particles changes with temperature because of the variation of repulsive coefficient $a_{ij}$. To satisfy the conservation of energy, the change of potential energy is considered to be balanced by a change of internal energy. Specifically, the total energy for each pair is considered invariable and its variation is zero upon time integration, i.e., $\Delta E_{ij}=\Delta U_{ij}+\Delta e_i - Q_i + \Delta e_j - Q_j=0$, where $\Delta e_i=C_v\Delta T_i$ and $\Delta e_j=C_v\Delta T_j$ are the changes of internal energy of particles $i$ and $j$, and $Q_i$ and $Q_j$ represent their net heat fluxes.

We consider an eDPD system containing one microgel bead in solution, and take LCST-type TRP as an example. Here, the results are interpreted in terms of the reduced DPD units, unless specified otherwise. For applications to specific materials, we refer interested readers to Refs. \citen{1997Groot,2014Z_Li} for parameterization of DPD systems. The microgel bead is made up of many cross-linked linear polymer chains (see Fig. S1, ESI$\dag$). Each polymer chain consists of 50 eDPD particles sequentially connected by harmonic springs, and cross-links with a density of approximately $3\%$ total bonds are randomly distributed in the microgel bead. The eDPD system of a TRP microgel bead surrounded by solvent particles is initialized at a low temperature $T_0=0.8T_c$. Simulations involving half a million particles are performed using a GPU-accelerated DPD \USERMESO package \cite{2013Tang}. Since the particle system is constructed with random initial configurations, we run the eDPD simulations at $T_0=0.8T_c$ for 100 time units to obtain the thermal equilibrium state. Then, similarly to the differential scanning calorimetry (DSC) experiments, the temperature of the eDPD system is increased linearly as a function of time, i.e., from $T_0=0.8T_c$ to $T_1=1.4T_c$ within 1500 time units.

\begin{figure}[b!]
  \centering
  \includegraphics[width=0.45\textwidth]{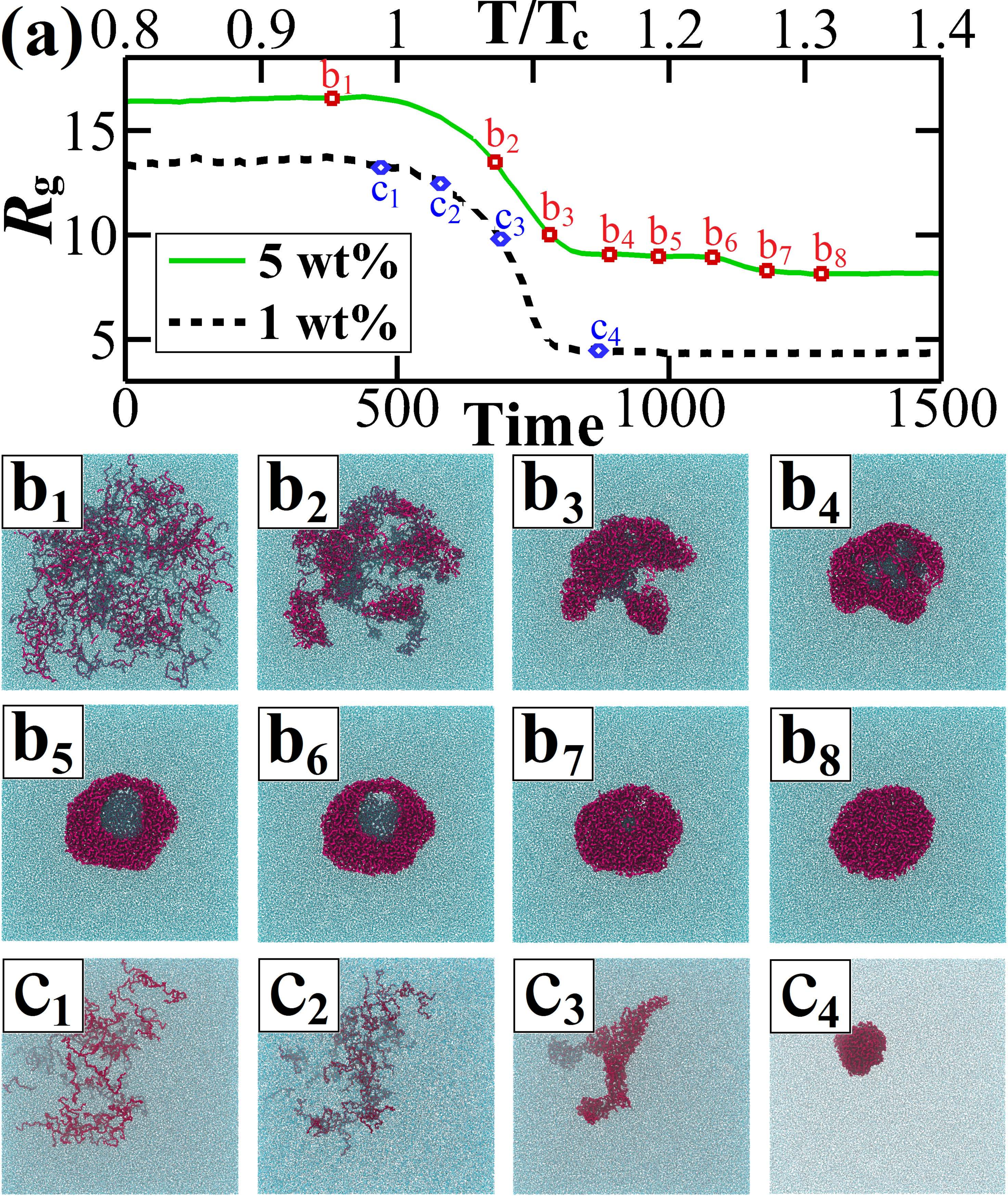}\\
  \caption{(a) Evolution of gyration radius $R_g$ of LCST-type thermoresponsive microgels with concentrations of $1~{\rm wt}\%$ and $5~{\rm wt}\%$ during heating. ($b_1$-$b_8$) and ($c_1$-$c_4$) show cross sections of their transient microstructure corresponding to the changes of $R_g$.}\label{fig:2}
\end{figure}

An obvious observation in the eDPD simulation is the configurational change and associated change in size of the TRP microgel bead. To quantify the deformation of the microgel bead during the heating process, we computed its instantaneous gyration radius $R_g$. Fig. \ref{fig:2}(a) shows the time evolution of $R_g$ during heating from $0.8T_c$ to $1.4T_c$, which contains a significant decrease of $R_g$ corresponding to a phase transition between $T=T_c$ and $T=1.1T_c$. More specifically, eight configurations of a large microgel bead ($5~{\rm wt\%}$) along the $R_g$ curve are presented in Figs. \ref{fig:2}(b$_1$-b$_8$), and four snapshots for a small microgel bead ($1~{\rm wt\%}$) are shown in Figs. \ref{fig:2}(c$_1$-c$_4$). The corresponding points are marked with symbols on the curve of $R_g$ shown in Fig. \ref{fig:2}(a). At temperatures below the critical temperature of phase transition, i.e., $T<T_c$, the TRPs are hydrophilic leading to a fully swollen state of the microgels, as displayed in Figs. \ref{fig:2}(b$_1$, c$_1$). A swollen microgel bead has the maximum volume and corresponds to the maximum gyration radius. Thus, the curves of $R_g$ have a plateau when $T<T_c$.  However, as the temperature increases, the hydrophilic-hydrophobic transition occurs near the critical temperature $T\approx T_c$, above which the TRPs become hydrophobic, and hence the microgel bead starts to collapse until it turns into a compact collapsed globule as shown in Figs. \ref{fig:2}(b$_8$, c$_4$).

For self-aggregation of TRP in the coil-to-globule phase transition process, two different mechanisms dependent on the size of TRP molecules dominate the dynamics \cite{1995Tiktopulo}. Small TRP molecules undergo an ``all-or-none" process while large TRP molecules behave as if they consist of quasi-independent ``domains" \cite{1995Tiktopulo}. Figs. \ref{fig:2}(b$_1$-b$_8$) and (c$_1$-c$_4$) provide dynamic details of the phase transition process. Our simulations confirm that a small microgel bead ($1~{\rm wt\%}$) has the ``all-or-none" coil-to-globule transition, which is a relatively simple process (Fig. \ref{fig:2}(c$_1$-c$_4$)). However, a larger microgel bead ($5~{\rm wt\%}$) has many ``independent domains" that start their self-aggregation processes simultaneously. In particular, porous structures can be observed at the beginning of coil-to-globule transition of the microgel bead, as shown in Fig. \ref{fig:2}(b$_2$). Those pores may trap solvents and then merge them inside the microgel bead (Fig. \ref{fig:2}(b$_4$-b$_6$)). In general, a big solvent droplet trapped inside a hydrophobic microgel bead is unstable. The droplet will finally escape, and then a compact collapsed globule is observed in Fig. \ref{fig:2}(b$_8$).

\begin{figure}[b!]
  \centering
  \includegraphics[width=0.4\textwidth]{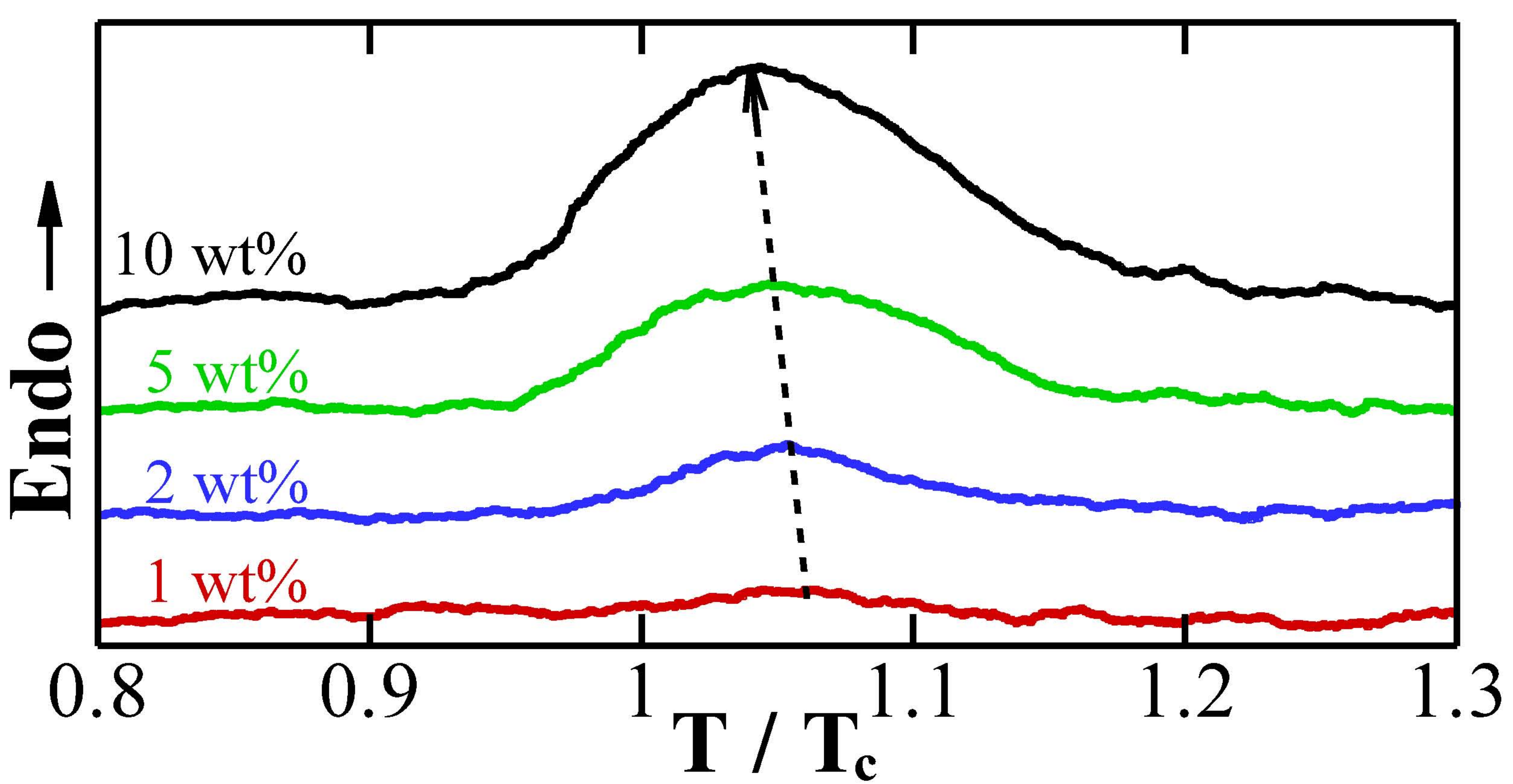}\\
  \caption{Thermogram analog to differential scanning calorimetry (DSC) heating curve with a fixed scanning rate at different concentrations ($1~{\rm wt}\%$, $2~{\rm wt}\%$, $5~{\rm wt}\%$, $10~{\rm wt}\%$).}\label{fig:3}
\end{figure}

Experiments \cite{1995Tiktopulo, 2012Sun} reported that LCST-type TRP undergo a coil-to-globule transition upon temperature increase, which is accompanied by cooperative heat absorption and results in an endothermic peak in the differential scanning calorimetry (DSC) thermogram. In particular, the endothermic peak depends on polymer concentration and shifts slightly to lower temperatures with concentrating polymer solutions \cite{2012Sun}. Figure \ref{fig:3} shows the thermal events that require energy flow to eDPD systems with a fixed scanning rate (i.e., $4.0\times10^{-4}T_c$ per time unit) for different concentrations ($1~{\rm wt}\%$, $2~{\rm wt}\%$, $5~{\rm wt}\%$, $10~{\rm wt}\%$). We observe a shift of the endothermic peak to lower temperatures as the concentration increases. Since the effects of ionic additives and PH are excluded in the present model, the only reason responsible for the shift is the difference of self-aggregation dynamics. At the beginning of the phase transition, the nucleation dynamics of TRP is reversible. For small microgels, the ``all-or-none" process needs more time to initialize the phase transition. However, many ``independent domains" of larger microgels start their nucleation processes simultaneously, which reduces the delay of phase transition. As a demonstration case of the proposed model, our simulations confirm that small TRP microgels in solution experience an ``all-or-none" process while large TRP microgels are dominated by a ``domain" mechanism \cite{1995Tiktopulo}(the independent formation of cooperative units). The different dynamic processes contribute to the shift of endothermic peak in  DSC thermograms.

In summary, we present a non-isothermal mesoscopic model that can be used to simulate thermally induced phase transition of TRP and to provide dynamic details of the phase transition process. Since the model conserves the energy of system in the simulations, it is also able to capture the underlying energy variation associated with the phase transition. This mesoscopic model is a promising candidate for modeling thermally induced phase transition of various thermoresponsive polymers and can assist $in\ silico$ design for engineering and biomedical applications of thermoresponsive materials.

This work was supported by the new DOE Center on Mathematics for Mesoscopic Modeling of Materials (CM4). Computations were performed at the Argonne Leadership Computing Facility through the INCITE program.

\footnotesize{
\bibliography{references}
\bibliographystyle{rsc} }

\end{document}